\documentstyle[cite,11pt]{article}
\newcommand{\be}{\begin{equation}}
\newcommand{\ee}{\end{equation}}
\newcommand{\beqn}{\begin{eqnarray}}
\newcommand{\eeqn}{\end{eqnarray}}
\newcommand{\beqnn}{\begin{eqnarray*}}
\newcommand{\eeqnn}{\end{eqnarray*}}

\def\a{\alpha}

\def\ep{\epsilon}

\begin{document}
%\draft

\title{Nonstationary Casimir effect in cavities with two resonantly
coupled modes}

\author{A. V. Dodonov and
V. V. Dodonov
\thanks{ e-mail: vdodonov@df.ufscar.br}
\thanks{%
on leave from Lebedev Physics Institute and Moscow Institute of Physics and
Technology, Russia}
%}
\\
%\address{
Departamento de F\'{\i}sica, Universidade Federal de
S\~ao Carlos,\\
Via Washington Luiz, km 235, 13565-905  S\~ao Carlos,  SP,  Brasil}
\date{}
\maketitle

%\small

\begin{abstract}
        We study the peculiarities of the nonstationary Casimir effect
(creation of photons in cavities with moving boundaries)
in the special case of two resonantly coupled modes with frequencies
$\omega_0$ and $(3+\Delta)\omega_0$, parametrically excited due to small
amplitude oscillations of the ideal cavity wall at the frequency
$2\omega_0(1+\delta)$ (with $|\delta|,|\Delta|\ll 1$).
The effects of thermally induced oscillations in time dependences of the
mean numbers of created photons and the exchange of quantum purities
between the modes are discovered. Squeezing and photon
distributions in each modes are calculated for initial vacuum and thermal
states. A possibility of compensation of detunings is shown.

\end{abstract}

\vspace{5mm}



\newpage %\twocolumn
\setlength{\baselineskip}{18pt}
%\renewcommand{\baselinestretch}{0.5}

%\narrowtext
%\twocolumn
%\begin{multicols}{2}

%\normalbaselines

\section{Introduction}

Classical and quantum phenomena in cavities with moving boundaries attracted
attention of many researchers for a long time (see review \cite{review}).
Especially popular this topic
became in the last decade, being known now under the names {\it %
nonstationary Casimir effect\/} \cite{DKM89},
{\it dynamical Casimir effect\/} \cite{Sch},
or {\it mirror (motion) induced radiation\/} \cite{BE,Lamb}.
One of several theoretical results obtained in the last years was the
prediction of the exponential growth of the energy of the field under the
resonance conditions, when the wall performs vibrations at the frequency
which is a multiple of the unperturbed field eigenfrequency
\cite{Lamb,Law,D95,DK96}.

In most papers, the special case of the one-dimensional
cavity with an equidistant spectrum of unperturbed field modes
was studied.
Then an infinite number of modes are excited due to an intermode
interaction (resulting from the Doppler effect on the moving boundary),
and the total number of created photons depends on time as $t^2$
(but the total energy increases exponentially)
\cite{review,DK96,D98,DA,AD00}.
A more realistic case of a three-dimensional cavity
was considered, e.g., in \cite{ViHaJa95,Mund,Ji98a}.
In particular, simple
analytical solutions describing the resonance creation of photons
in cavities with totally nonequidistant spectra,
when only one mode is in resonance with the vibrating wall,
were given in \cite{review,D95,DK96}.
However, the case of a few resonantly interacting modes is also of a
great interest. In particular, the interaction between the excited field
mode and the detector, approximated either by a harmonic oscillator
or by a two-level atom, was considered in
\cite{D95,DK96,Law95,JaVi96,TaKo98,Janow98,Fedot00}.

It was shown recently \cite{Maz01}, that
for rectangular three- and two-dimensional cavities,
there exist special configurations,
when two (or more) modes can occur in resonance with the moving wall and
between themselves.
For example, the eigenmode spectrum in a cubical cavity of length $L$
is given by the formula $\omega_{klm}=(\pi c/L)\sqrt{k^2 +l^2 +m^2}$.
Since $\omega_{511} =3\omega_{111}$, both the modes, $\{111\}$ and $\{511\}$,
will be excited if the wall oscillates at the frequency $2\omega_{111}$.
In such a case, it appears \cite{Maz01} that the energies of both the modes
increase in time exponentially (with some additional oscillations), but
the increment of the exponential growth is twice smaller than in the case
of a single resonance mode.

Our goal is to give analytical solutions describing the case
of two resonantly interacting modes in a more or less generic situation,
without specializing the form of the cavity.
We take into account the
possibility of detuning between the frequency of the vibrating wall and
the twice fundamental field eigenfrequency, and detuning between the
frequency of the second mode and the triple fundamental eigenfrequency
(however, we do not consider the effects of damping).
Using the method of slowly varying amplitudes, we obtain explicit solutions
of equations of motion
for the canonical operators describing the fields in coupled modes,
which enable us to calculate the mean energy and the invariant uncertainty
product for each mode.
We analyze the evolution of the photon distribution functions
for each mode and show the possibility of strong {\em squeezing\/} in the
long-time limit (for other configurations this was done in
\cite{D95,DK96,DA}).

\section{General solution for two coupled modes}

We use the Hamiltonian approach proposed by Law \cite{Law} and
developed in \cite{Plun} (for other references see \cite{review}).
Consider the scalar massless field $\Phi({\bf r},t)$, satisfying
the wave equation $\Phi_{tt}=\nabla^2\Phi$ inside the cavity and the
Dirichlet boundary condition $\Phi=0$ on the boundary.
We assume that we know the complete orthonormalized set of eigenfunctions
(and eigenfrequencies) of the Laplace equation
$\nabla^2 f_{\a}({\bf r}) +\omega_{\a}^2 f_{\a}({\bf r})=0$
in the case of stationary cavity.
Now suppose that a part of the boundary is a plane surface moving
according to a {\em prescribed\/} law of motion $L(t)$ (for the most recent
study of the case when $L(t)$ is a {\em dynamical variable\/} due to
the back reaction of the field see \cite{Cole01}).
Expanding the field $\Phi({\bf r},t)$ over ``instantaneous''
eigenfunctions $f_{\a}({\bf r};L(t))$,
\be
\Phi({\bf r},t)=\sum_{\a} q_{\a}(t) f_{\a}({\bf r};L(t)),
\label{expansion}
\ee
we satisfy automatically the boundary conditions. Then the dynamics of the
field is described completely by the dynamics of the generalized
coordinates $q_{\a}(t)$, which, in turn,
can be derived from the
{\em time-dependent Hamiltonian\/} \cite{Plun}
\be
H(t) = \frac12\sum_{\a}\left[p_{\a}^2 +\omega_{\a}^2(L(t)) q_{\a}^2\right]
+\frac{\dot{L}(t)}{L(t)}\sum_{\a \neq \beta} p_{\a} m_{\a\beta}q_{\beta}
\label{genHam}
\ee
with antisymmetrical time-independent coefficients
\be
m_{\a\beta}= - m_{\beta\a} =L \int dV
\frac{\partial f_{\a}({\bf r};L)}{\partial L}f_{\beta}({\bf r};L).
\label{defmab}
\ee
For example, in the case of a
rectangular three-dimensional cavity, the eigenmodes are the well known
products of three sine functions like $\sin\left(\pi k_x x/L_x\right)$,
labeled by three
natural numbers $k_x,k_y,k_z$. If one surface of the parallelepiped,
perpendicular to the $x$-axis, moves in the $x$-direction
(so that the $L_x$-dimension of the cavity is a function of time), then
\cite{Maz01}
\be
m_{{\bf k}{\bf j}} = (-1)^{k_x+j_x} \frac{2 k_x j_x}{j_x^2 - k_x^2}
\delta_{k_y j_y}\delta_{k_z j_z}.
\label{m3D}
\ee

We are interested in the case when one of the cavity walls performs
small oscillations with the frequency $\Omega$ close to the double frequency
of some unperturbed mode $\omega_1^{(0)} \equiv 1$ (i.e., we normalize all
frequencies by $\omega_1^{(0)}$), so that the time-dependent frequency
$\omega_1(t)$ reads
\be
\omega_1(t) = 1 + 2\ep \cos(2\overline{\omega}t),
\quad \overline{\omega}=1 +\delta,
\label{Lt}
\ee
where we assume that $|\delta|\ll 1$ and $|\ep|\ll 1$.
Also we suppose that the unperturbed field frequency spectrum includes
the frequency $\omega_3^{(0)} =3 +\Delta$ with $|\Delta|\ll 1$,
but it does not
contain frequencies close to $5\omega_1^{(0)}$. Then we have the case of two
resonantly interacting modes, and it is sufficient to consider only the
part of the total Hamiltonian (\ref{genHam}) related to these modes:
\beqn
H_{13} &=&\frac12\left(p_1^2 + p_3^2\right) + \frac12
\left[1 + 4\ep \cos(2\overline{\omega}t)\right] x_1^2
\nonumber \\ &&
+\frac12 \left[9+6\Delta +\tilde{\ep} \cos(2\overline{\omega}t)\right]x_3^2
\nonumber \\ &&
+3\mu\ep\sin(2\overline{\omega}t)\left(p_1 x_3 -p_3 x_1\right).
\label{H2}
\eeqn
The constant parameter $\mu$
is proportional to the coefficient $m_{12}$ in (\ref{genHam}).
Writing (\ref{H2}) we have neglected the
second order terms with respect to $\ep$ and $\Delta$.
Since the time-dependent frequency shift
$\omega_1(t)-\omega_1^{(0)} \sim L(t)-L_0$ is chosen
to be proportional to $\cos(2\overline{\omega}t)$, the last term in
(\ref{H2}), which is proportional to $\dot{L}(t)$, must depend on time
as $\sin(2\overline{\omega}t)$.
All numerical coefficients
in (\ref{H2}) are chosen in order to avoid an appearance of fractions in
formulas below.
Parameter $\tilde{\ep}$
%in the time-dependent frequency squared $\omega_3^2(t)$
has the same order of magnitude as $\ep$,
but it does not affect the solution in the
zeroth order approximation (in which we are interested here),
as will be shown below.

Hamiltonian (\ref{H2}) results in the following differential
equations for the generalized coordinates $x_1$ and $x_3$
(we neglect corrections of the second order):
\beqn
\ddot{x}_1 &=& -\left[1 + 4\ep \cos(2\overline{\omega}t)\right] x_1
\nonumber \\ && +
24\mu\ep\left[ \cos(2\overline{\omega}t) x_3
+ \sin(2\overline{\omega}t) \dot{x}_3\right],
\label{ddotx1}
\eeqn
\beqn
\ddot{x}_3 &=&
- \left[9+6\Delta +\tilde{\ep} \cos(2\overline{\omega}t)\right] x_3
\nonumber \\ && -
24\mu\ep\left[ \cos(2\overline{\omega}t) x_1
+ \sin(2\overline{\omega}t) \dot{x}_1\right].
\label{ddotx2}
\eeqn
We solve equations (\ref{ddotx1}) and (\ref{ddotx2}), using the method
of slowly varying amplitudes \cite{Louis,Land,Bogol})
(which was applied earlier in studies
\cite{D95,DK96,D98,D98a}).
Namely, we look for the solutions in the form
\be
x_{k}(t)= \xi_{k}^{+}(t) e^{ik\overline{\omega}t}
+\xi_{k}^{-}(t) e^{-ik\overline{\omega}t}, \quad k=1,3
\label{formx1}
\ee
where each coefficient $\xi_{k}^{\pm}$ is a {\em slowly varying function
of time\/}, whose derivatives are proportional to the small parameters
$\ep,\delta,\Delta$, so that we can neglect the second-order derivatives
$\ddot\xi_{k}^{\pm}$. Then, equating the coefficients at
$\exp[\pm i\overline{\omega}t]$ in equation (\ref{ddotx1}) and
coefficients at $\exp[\pm 3i\overline{\omega}t]$ in equation (\ref{ddotx2})
(and neglecting the terms proportional to squares and products of small
coefficients $\ep,\delta,\Delta$),
we obtain a set of equations with {\em constant\/} coefficients
for the slowly varying amplitudes. It is convenient to write this set in
the matrix form $d{\bf v}/dt={\cal A}{\bf v}$, introducing vector
${\bf v}=\left(\xi_{1}^{+},\xi_{1}^{-},\xi_{3}^{+},\xi_{3}^{-}\right)$
and matrix
\be
{\cal A}=\left\Vert
\begin{array}{cccc}
-i\delta & i\ep      & 12i\mu\ep           & 0                   \\
-i\ep    & i\delta   &  0                  & -12i\mu\ep           \\
4i\mu\ep & 0         & i(\Delta - 3\delta) & 0                     \\
0        & -4i\mu\ep &  0                  & -i(\Delta - 3\delta)
\end{array}
\right\Vert .
\label{matrA}
\ee
Strictly speaking, function $x_1(t)$ contains also terms proportional
to $\exp[\pm 3i\overline{\omega}t]$, as well as
function $x_3(t)$ contains terms proportional
to $\exp[\pm i\overline{\omega}t]$. However,
the amplitudes of these additional terms are
proportional to the small parameters $\ep,\delta,\Delta$, so we neglect
them, confining ourselves to the principal terms whose amplitudes are
of the zeroth order with respect to small parameters. Just for this
reason, the parameter $\tilde\ep$ does not appear in matrix ${\cal A}$:
while the term $4\ep \cos(2\overline{\omega}t)$ standing at $x_1$ in
equation (\ref{ddotx1}) connects the slowly varying amplitudes
$\xi_{1}^{+}$ and $\xi_{1}^{-}$, a similar term
$\tilde\ep \cos(2\overline{\omega}t)$ standing at $x_3$ in (\ref{ddotx2})
does not connect the amplitudes $\xi_{3}^{+}$ and $\xi_{3}^{-}$
between themselves,
but it gives small corrections to the amplitudes of higher-order
terms in $x_3$, oscillating with frequencies $\overline{\omega}$ and
$5\overline{\omega}$. A more strict (from the mathematical
point of view) method, based on the multiple scale analysis, was
exposed in \cite{Maz01}. In the zeroth-order approximation it leads to
the same results.

Matrix ${\cal A}$ has four eigenvalues, $\pm\lambda_{\pm}$, where
\beqn
\lambda_{\pm} &=&\frac1{\sqrt2}\sqrt{a \pm \sqrt{c}}
\label{lam1}\\ &\equiv&
\frac12\left(\sqrt{a+\sqrt{b}} \pm \sqrt{a-\sqrt{b}}\right),
\label{lam2}
\eeqn
\be
a= \ep^2(1-\nu) -\delta^2 -(\Delta - 3\delta)^2,
\label{def-a}
\ee
\beqn
b&=&\left[2(\delta-\ep)(\Delta-3\delta)+\nu\ep^2\right]
\nonumber\\&\times&
\left[2(\delta+\ep)(\Delta-3\delta)+\nu\ep^2\right],
\label{def-b}
\eeqn
\beqn
c&\equiv& a^2-b = 2\ep^2 \nu\left[(\Delta - 4\delta)^2 -\ep^2\right]
\nonumber \\ &&
+\left[\ep^2 + (\Delta - 2\delta)(\Delta - 4\delta)\right]^2,
\label{def-c}
\eeqn
\be
\nu \equiv 96\mu^2 .
\label{def-nu}
\ee
Comparing our matrix ${\cal A}$ (\ref{matrA}) with a similar matrix
found in \cite{Maz01} for the rectangular cavity, one can verify that
if modes $\{k_x,m,n\}$ and $\{j_x,m,n\}$ are in resonance, then
$\mu=j_x/(12k_x)$ (different phases of elements of matrices in \cite{Maz01}
and here are due to different choices of trigonometrical dependences
in function $L(t)$: we use $\cos$-function instead of $\sin$-function
in \cite{Maz01}). In particular, for the modes $\{111\}$ and $\{511\}$
of the cubical cavity we have $\nu=50/3$. Due to this explicit example
(and other examples related to the rectangular cavities), we assume
hereafter that parameter $\nu$ is large, so that it satisfies
the inequality $\nu \gg 1 $.

\section{Exact resonance}

We start with the simplest case of the exact
resonance, $\delta=\Delta=0$. Then formula (\ref{lam2}) yields
\be
\lambda_{\pm}^{(\nu)}=\frac{\ep}{2}\left(1 \pm \sqrt{1-2\nu}\right).
\label{lam0}
\ee
If $\nu=0$ (no intermode coupling), then we obtain the increment of the
exponential growth $\lambda_+^{(0)}=\ep$, in accordance with the solutions
of the single-mode problem found in \cite{D95,DK96,D98a}.
On the contrary, for $\nu > 1/2$ the increment
(real part of $\lambda_{\pm}$) does not depend on the form of the cavity
(which is ``hidden'' in the value of $\nu$), being
exactly twice smaller than in the single-mode case.
In the special case of rectangular cavities this result was obtained
in \cite{Maz01}.

%Calculating eigenvectors of matrix (\ref{matrA}),
After some straightforward calculations we arrive at the following
explicit expressions for the time dependences of
the generalized coordinates
%and the canonically conjugated momenta
(as far as we neglect all corrections of the order of $\ep$ in the amplitude
coefficients, we may identify the canonical momenta with velocities,
$p_k=\dot{x}_k$, $k=1,2$):
\beqn
&&x_1(t) = x_1(0)\left[C_{1}^{-}\,\cos(\rho\tau)
+S_{1}^{-}\,\frac{\sin(\rho\tau)}{\rho}\right]
\nonumber\\&&
- p_1(0)\left[S_{1}^{-}\,\cos(\rho\tau)
+C_{1}^{-}\,\frac{\sin(\rho\tau)}{\rho}\right]
\nonumber\\&&
+8\mu\frac{\sin(\rho\tau)}{\rho}
\left[3S_{1}^{-}\, x_3(0) +C_{1}^{-}\, p_3(0)\right],
\label{x1t}
\eeqn
\beqn
&&x_3(t) = x_3(0)\left[C_{3}^{+}\,\cos(\rho\tau)
-S_{3}^{+}\,\frac{\sin(\rho\tau)}{\rho}\right]
\nonumber\\&&
+\frac13 p_3(0)\left[S_{3}^{+}\,\cos(\rho\tau)
-C_{3}^{+}\,\frac{\sin(\rho\tau)}{\rho}\right]
\nonumber\\&&
-8\mu\frac{\sin(\rho\tau)}{\rho}
\left[S_{3}^{+}\, x_1(0) -C_{3}^{+}\, p_1(0)\right],
\label{x2t}
\eeqn
where
\be
\tau \equiv \frac12 \ep t, \quad \rho=\sqrt{2\nu-1},
\label{deftaurho}
\ee
\beqnn
&&C_{k}^{\pm}(\tau;t) = \cosh\tau \cos(k\overline{\omega}t)
\pm \sinh\tau \sin(k\overline{\omega}t),
%\label{defCpm}
\\
&&S_{k}^{\pm}(\tau;t) = \sinh\tau \cos(k\overline{\omega}t)
\pm \cosh\tau \sin(k\overline{\omega}t).
%\label{defSpm}
\eeqnn
Similar expressions for the time dependent canonical momenta can be
easily obtained by differentiating equations (\ref{x1t})-(\ref{x2t})
with respect to
the ``fast time'' $t$, considering the ``slow time'' variable $\tau$ as
an independent parameter and neglecting corrections of the order of
$\ep,\delta,\Delta$ in the amplitude coefficients. The following relations
hold in this approximation:
\[
\frac{\partial C_{k}^{\pm}}{\partial t}= \pm k S_{k}^{\mp}, \quad
\frac{\partial S_{k}^{\pm}}{\partial t}= \pm k C_{k}^{\mp}.
\]
Symbols $x_k,p_k$ in equations (\ref{x1t})-(\ref{x2t}) can be considered
both as classical variables and quantum operators in the Heisenberg
picture, due to the linearity of the problem (or due to the quadratic
nature of Hamiltonian (\ref{genHam})). Using equations
(\ref{x1t})-(\ref{x2t}) and their momentum counterparts,
one can calculate mean values of squares and
products of canonical variables (operators) at any moment of time,
provided such mean values were known at the initial moment $t=0$.
We confine ourselves to the simplest case, when initially the field
modes were in thermal states with the mean photon numbers
$(\theta_1-1)/2$ and $(\theta_3-1)/2$, where $\theta_k=\coth(k\beta/2)$,
$\beta$ being inverse absolute temperature in dimensionless units.
One can check the relations
\[
\theta_{31}\equiv \frac{\theta_{3}}{\theta_{1}} =\theta_{13}^{-1}
=\frac{\theta_{1}^2 +3}{3\theta_{1}^2 +1}, \quad
1\ge \theta_{31} \ge \frac13.
\]
The mean energies in each mode,
${\cal E}_k =\frac12\langle p_k^2 + \omega_k^2 x_k^2\rangle$,
depend on time as follows,
\beqn
{\cal E}_1 &=& \frac{\theta_1}{2}\Bigg\{
\cosh(2\tau)\left[
\frac{\sin^2(\rho\tau)}{\rho^2}
\left(1 +2\nu\,\theta_{31}\right)\right.
\nonumber\\ && \left.
+\cos^2(\rho\tau)\right]
+ \sinh(2\tau) \frac{\sin(2\rho\tau)}{\rho}\Bigg\},
\label{E1t}
\eeqn
\beqn
{\cal E}_3 &=& \frac{3\theta_3}{2}\Bigg\{
\cosh(2\tau)\left[
\frac{\sin^2(\rho\tau)}{\rho^2}
\left(1 +2\nu\,\theta_{13}\right) \right.
\nonumber\\ && \left.
+\cos^2(\rho\tau)\right]
- \sinh(2\tau) \frac{\sin(2\rho\tau)}{\rho}\Bigg\}.
\label{E2t}
\eeqn
For rectangular cavities (and, perhaps, for others),
$\nu\gg 1$ in the case of intermode resonance.
Then $\rho^2\approx 2\nu \gg 1$, so that equations (\ref{E1t})
and (\ref{E2t}) can be simplified:
\beqnn
{\cal E}_1 &\approx& \frac12 \cosh(2\tau)
\left[\theta_1\cos^2(\rho\tau) +\theta_3 \sin^2(\rho\tau)\right],
%\label{E1app}
%\ee
%\be
\\
{\cal E}_3 &\approx& \frac32 \cosh(2\tau)
\left[\theta_3\cos^2(\rho\tau) +\theta_1 \sin^2(\rho\tau)\right].
%\label{E2app}
\eeqnn
For the initial vacuum states ($\theta_1=\theta_3=1$) we have
a monotonous growth of energy and number of photons in each mode:
${\cal E}_k\approx (\omega_k^{(0)}/2)\cosh(2\tau)$, $k=1,2$.
On the contrary, for high-temperature initial states (with equal
temperatures), $\theta_1=3\theta_3 \gg 1$, which results in
strong oscillations of mean energies and numbers of photons:
\beqnn
{\cal E}_1 &\approx& \frac{\theta_1}{2} \cosh(2\tau)
\left[ 1 -\frac23 \sin^2(\rho\tau)\right],
\\
{\cal E}_3 &\approx& \frac{\theta_1}{2} \cosh(2\tau)
\left[ 1 +2 \sin^2(\rho\tau)\right].
\eeqnn
In Fig.~1 we show normalized time dependences of the mean energies
(calculated with the aid of complete formulas (\ref{E1t}) and (\ref{E2t}))
in each resonant mode in the low- and high-temperature limits.
Note that one has $\theta_1 \approx 140$,
if $L_0=1\,$cm and $T=300\,$K.

Since Hamiltonian (\ref{genHam}) is {\em quadratic\/} with respect to
canonical coordinates and momenta), the initial thermal state
is transformed with time to a generic {\em Gaussian quantum state\/}
(whose density matrix or Wigner function is a Gaussian exponential).
The single-mode density matrices of such states are completely characterized
(in the case of the mean values of quadrature components), besides the
mean energy, by the {\em invariant uncertainty product\/} (IUP)
\be
{\cal D}\equiv \langle x^2\rangle \langle p^2\rangle
-\langle (xp+px)/2\rangle^2.
\label{IUP}
\ee
For $\nu>1/2$ this quantity varies in time periodically. For the first mode,
\beqn
{\cal D}_1 &=& \frac{\theta_1^2}{4}\left[\cos^4(\rho\tau) +
\sin^2(2\rho\tau)\frac{2\nu\,\theta_{31}-1}{2(2\nu-1)}
\right.\nonumber \\&& \left. +
\sin^4(\rho\tau)\left(\frac{2\nu\,\theta_{31}+1}{2\nu-1}
\right)^2\right].
\label{Dtau}
\eeqn
For another excited mode
one should interchange indices $1$ and $3$ in (\ref{Dtau}).

The {\em purity\/} of quantum Gaussian states is expressed through IUP as
$\mbox{Tr}(\hat\rho^2)=(4{\cal D})^{-1/2}$ (here $\hat\rho$ is the
statistical operator of the state).
For initial vacuum states,
\be
{\cal D}_{1,3}=
\frac14\left[1 + \frac{8\nu}{(2\nu-1)^2}\sin^4(\rho\tau)\right],
\label{Dvac}
\ee
so that the states remain practically pure, if $\nu \gg 1$.
This is a significant difference from the case of a single resonance
mode coupled to a harmonic oscillator detector, studied in \cite{D95,DK96}.
If $\theta_1 \neq \theta_3$ (nonzero initial temperature) and $\nu \gg 1$,
then
\[
{\cal D}_1 \approx \frac14\left[ \theta_1 \cos^2(\rho\tau)+
\theta_3 \sin^2(\rho\tau)\right]^2,
\]
so that coupled modes {\em exchange their purities\/} at the moments
when $\sin(\rho\tau)=1$.

The {\em squeezing coefficient\/}, defined as the ratio of the minimal
value of any canonical variance ($\sigma_x$ or $\sigma_p$) for the
period of fast oscillations (with frequencies $\overline{\omega}$ or
$3\overline{\omega}$) to the vacuum value ($(2\omega_k^{(0)})^{-1}$ or
$\omega_k^{(0)}/2$), can be expressed through
$\tilde{\cal E}\equiv {\cal E}/\omega_k^{(0)}$ and ${\cal D}$ as
\cite{princ,Pavel}
\be
s=\frac{2{\cal D}}{\tilde{\cal E} +\sqrt{\tilde{\cal E}^2 -{\cal D}}}.
\label{s}
\ee
For thermal states, $s(0)=\theta$, but
asymptotically one can obtain any desired degree of squeezing
(even for initial high-temperature states) in each resonant mode,
because $s \approx {\cal D}(\tau)/\tilde{\cal E}(\tau)$ for $\tau \gg 1$.
Evidently, $\tilde{\cal E}$ is nothing but the mean number of photons
in the mode, if $\tilde{\cal E} \gg 1$. The variance of the photon
distribution in the Gaussian states is given by the formula \cite{1mod}
\[
\sigma_n= 2 \tilde{\cal E}^2 - {\cal D} -1/4.
\]
The photon statistics is strongly super-Poissonian for
$\tau \gg 1$, when
$\sigma_n/\langle n\rangle \approx 2 \tilde{\cal E}\gg 1$.
Nonetheless, the quantum states of each excited mode become highly
nonclassical for $\tau\gg 1$.
It is seen from the photon distribution function (PDF),
which can be expressed in terms of two parameters,
$\tilde{\cal E}$ and ${\cal D}$,
for any Gaussian state
with zero mean values of quadrature variables \cite{1mod}):
\begin{eqnarray}
{\cal P}_n&=&\frac{2}{\sqrt{1+4\tilde{\cal E}+4{\cal D}}}
\left(\frac{1+4{\cal D} -4\tilde{\cal E}}{1+4{\cal D}+4\tilde{\cal E}}
\right)^{n/2}\nonumber\\
&\times&P_n\left(\frac{4{\cal D}-1}
{\sqrt{(4{\cal D}+1)^2-16\tilde{\cal E}^2}}\right).
\label{dist}
\end{eqnarray}
Here $P_n(z)$ is the Legendre polynomial.
Initially $\tilde{\cal E}^2(0) ={\cal D}(0)$, and (\ref{dist})
is a monotonous geometric distribution,
${\cal P}_n(0)= 2(\theta-1)^n/(\theta+1)^{n+1}$.
Since we are interested in the cases of large numbers of photons
created due to the parametric resonance,
it is convenient to use asymptotical forms of the exact distribution
(\ref{dist}) for $n\gg 1$. Note that the argument of the Legendre
polynomial in (\ref{dist}) is always outside the interval $(-1,1)$,
being equal to $1$ only at the initial moment (for thermal states).
With the course of time this argument
increases to $\infty$, becomes pure imaginary, and asymptotically
goes to zero, when $\tilde{\cal E}\gg {\cal D}$.
Therefore it is convenient to use the asymptotical formula for $n\gg 1$
\cite{Olver}
\begin{equation}
P_n(\cosh\xi )\approx
\left(\frac {\xi}{\sinh\xi}\right)^{1/2}
I_0\left(\left[n+1/2\right]\xi\right)
\label{as-Olv}
\end{equation}
(where $I_0(z)$ is the modified Bessel function), because it
holds even for complex values of variable $\xi$, provided
$\mbox{Re}\xi \ge 0$ and $|\mbox{Im}\xi |<\pi$. We are interested in the
case when the mean energy of each mode significantly exceeds its initial
value, i.e., $\tilde{\cal E} \gg \sqrt{{\cal D}}$. If ${\cal D} \gg 1$
(high temperature initial states), then it is possible that
$\tilde{\cal E} <{\cal D} +1/4$. In such a case, $\xi$ is real and large,
so that one can replace the modified Bessel function $I_0(x)$ by
its asymptotical expression for $x \gg 1$,
$I_0(x) \approx (2\pi x)^{-1/2}\exp(x)$.
Moreover, one can use either of approximate equalities
$\exp(x)\approx 2\cosh(x) \approx 2\sinh(x)$.

When $\tilde{\cal E} > {\cal D} +1/4$ (this is just the regime when
{\em squeezing\/} happens), the argument of the Legendre
polynomial becomes pure imaginary, and the variable $\xi$ becomes complex:
$\xi =-i\pi/2 +y$, with $y>0$. Then one should replace the modified
Bessel function in (\ref{as-Olv}) by the usual Bessel function
$J_0\left(\left[n+1/2\right][\pi/2 +iy]\right)$. Since the absolute value
of the argument of this function is also large, one can replace
$J_0(x)$ by $(\pi x/2)^{-1/2}\cos(x -\pi/4)$. In this case, we have
different expressions for even and odd values of index $n$
(this is especially clear from the initial formula (\ref{dist}), when the
argument of the Legendre polynomial is close to zero for
$\tilde{\cal E} \gg {\cal D}$).
Finally, we arrive at the following asymptotical expression,
which holds both for real and imaginary values of the argument
of the Legendre polynomial in (\ref{dist})
(provided $\tilde{\cal E}^2 \gg {\cal D}$):
\beqn
{\cal P}_n &\approx& \frac{\sqrt2
\left|1+4{\cal D} -4\tilde{\cal E}\right|^{n/2}}
{\sqrt{\pi n \tilde{\cal E}}
\left(1+4{\cal D} +4\tilde{\cal E}\right)^{(n+1)/2}}
\nonumber\\&&\times
 \left\{
\begin{array}{ll}
\cosh\left([n+1/2]\ln|\chi|\right),
& n \;\; \mbox{even}\\
\sinh \left([n+1/2]\ln|\chi|\right),
& n \;\; \mbox{odd}
\end{array}
\right.
\label{as-unif}
\eeqn
where
\be
\chi=\frac{4{\cal D}-1 +4\sqrt{\tilde{\cal E}^2 - {\cal D}}}
{\sqrt{(4{\cal D}+1)^2-16\tilde{\cal E}^2}}.
\label{defchi}
\ee
If the ratio $\tilde{\cal E} / {\cal D}$ is not very large, then
$\ln|\chi|$ is not small, and one can replace $\cosh$ and $\sinh$
functions in (\ref{as-unif}) by exponentials. In this case we have
a nonoscillating ``quasigeometric'' distribution, i.e.,
a geometric distribution modified by a slowly decreasing factor $n^{-1/2}$:
\be
{\cal P}_n \approx \frac{1}{\sqrt{2\pi n \tilde{\cal E}}}
\left(\frac{4{\cal D} -1 +4\sqrt{\tilde{\cal E}^2 - {\cal D}}}
{4{\cal D} +1 +4\tilde{\cal E}}\right)^{n+1/2}.
\label{dist-mon}
\ee
On the contrary, if $\tilde{\cal E} \gg {\cal D}$ (then $\chi$ is pure
imaginary and small), we have a strongly oscillating distribution
for $n < \tilde{\cal E} / {\cal D}$
(typical for squeezed states):
\beqn
{\cal P}_n &\approx& \sqrt{\frac{2}{\pi n \tilde{\cal E}}}
\left(1- \frac{4{\cal D} +1}{2\tilde{\cal E}}\right)^{n/2}
\nonumber\\ && \times
 \left\{
\begin{array}{ll}
\cosh\left(n \frac{4{\cal D} -1}{4\tilde{\cal E}}\right),
& n \;\; \mbox{even}\\
\sinh\left(n \frac{4{\cal D} -1}{4\tilde{\cal E}}\right)
& n \;\; \mbox{odd}
\end{array}
\right.
\label{asPn}
\eeqn
Only the ``tail'' of distribution (\ref{asPn}) does not oscillate
(and does not depend on ${\cal D}$):
\be
{\cal P}_n \approx \frac{\exp\left[-n/(2\tilde{\cal E})\right]}
{\sqrt{2\pi n \tilde{\cal E}}}, \quad
n \frac{4{\cal D} -1}{4\tilde{\cal E}} \gg 1.
\label{tail}
\ee
For the vacuum initial state, the PDF
oscillates from the beginning. For $\tau\gg 1$ and $\nu\gg 1$
we can write (for each excited mode)
\beqnn
{\cal P}_n &\approx& e^{-\tau}\sqrt{\frac{8}{\pi n}}
\left(1-4e^{-2\tau}\right)^{n/2}
\\ && \times
 \left\{
\begin{array}{ll}
\cosh\left( \frac{2n}{\nu}\sin^4(\rho\tau) e^{-2\tau}\right),
& n \;\; \mbox{even}\\
\sinh\left( \frac{2n}{\nu}\sin^4(\rho\tau) e^{-2\tau}\right),
& n \;\; \mbox{odd}
\end{array}
\right.
\eeqnn
Oscillations exist for any $n$ at the moments of time when
$\sin(\rho\tau)=0$.
Two examples of the photon distribution functions
for the initial vacuum and thermal states
(calculated using the exact formula (\ref{dist}))
are given in figures 2 and 3.

\section{Influence of detunings}

Now let us consider the generic case of nonzero detunings.
Hereafter we use the normalized parameters
$\tilde{\delta}\equiv \delta/\ep$ and $\tilde{\Delta}\equiv \Delta/\ep$.
The condition of the parametric resonance is that at least one of the
eigenvalues $\lambda_{\pm}$, given by equations
(\ref{lam1}) or (\ref{lam2}), has nonzero real part. Since we assume
that $\nu>1$, the coefficient $a$ (\ref{def-a}) is always negative.
Looking at the eigenvalues in the form (\ref{lam1}), we see that there are
two possibilities.

The first one is to choose $c<0$. Then complex number $a+\sqrt{c}$
has nonzero imaginary part, so its square root inevitably
has nonzero real part.
Designating $\tilde{\Delta}- 4\tilde{\delta}=\eta$, we see that $c<0$
for $|\eta| <\eta_c$, where the critical value $\eta_c$ (which
depends on $\nu$ and $\tilde{\delta}$) must be obviously less than $1$.
If $\nu\gg 1$ and $|\delta|\sim 1$, then $\eta_c$
is close to $1$, with corrections of the order of $\nu^{-1}$.
If $|\tilde{\delta}| \gg 1$, then for $1\sim|\eta|\ll|\tilde{\delta}|$
one can write
$c\approx 2\nu\left(\eta^2-1\right) + 4\tilde{\delta}^2\eta^2$
(taking into account that
$\tilde{\Delta}- 2\tilde{\delta} = 2\tilde{\delta} +\eta
\approx 2\tilde{\delta}$). In this way we obtain the following approximate
inequality, describing the region of resonance in the parameter plane
$\tilde{\delta},\tilde{\Delta}$ (see figure 4):
\be
|\tilde{\Delta}- 4\tilde{\delta}| < \eta_c\approx
\sqrt{\frac{\nu}{\nu +2\tilde{\delta}^2}}.
\label{cond-}
\ee
Consequently, one can always excite both the modes, compensating
one detuning by another.
For example, if $\tilde{\Delta}= 4\tilde{\delta}$
and $\nu \gg 1$, then
$b=a^2-c$ with $|c|\ll a^2$, so that
$\sqrt{b} \approx |a|-c/(2|a|)$.
In this case equation (\ref{lam2}) yields
\be
\lambda_{\pm} \approx \frac{\ep}{2}\left[\sqrt{\frac{\nu}
{\nu +2\tilde{\delta}^2}} \pm
i\sqrt{2\left(\nu+2\tilde{\delta}^2\right)}\right].
\label{Relam}
\ee
As far as $\tilde{\delta}^2 \ll \nu$, the maximal value of the increment
coefficient $\mbox{Re}(\lambda_{\pm})$ is practically the same as in the case
of strict resonance (\ref{lam0}).
Energies of both modes have the same order of magnitude (as in the
strict resonance case), therefore
this regime of excitation can be named ``symmetrical''.
However, with increase of $|\tilde{\delta}|$ the increment of energy growth
decreases (by the same law as the resonance width $\eta_c$), and for
$\tilde{\delta}^2\gg \nu$ we have
$ \mbox{Re}(\lambda_{\pm}) \approx \ep
\left[\nu/(8\tilde{\delta}^2)\right]^{1/2} \ll \ep
$.

The second possibility to swing the modes is to choose $b<0$ in
(\ref{def-b}). Then $c > a^2$, and the eigenvalue $\lambda_+$ is real
(whereas $\lambda_-$ is pure imaginary). There are two symmetrical
regions of swinging in the
parameter plane $\tilde{\delta},\tilde{\Delta}$, which are located
between the branches of hyperbolas (see figure 4):
\beqn
\frac{\nu}{2(1-\tilde{\delta})} < \gamma
< -\,\frac{\nu}{2(1+\tilde{\delta})}, &&  |\tilde{\delta}| >1,
\label{cond-2} \\
\frac{\nu}{2(1-\tilde{\delta})} < \gamma
\;\; \mbox{or} \;\;
\gamma < -\,\frac{\nu}{2(1+\tilde{\delta})},
&&  |\tilde{\delta}| \le 1,
\label{cond-2a}
\eeqn
where $\gamma \equiv \tilde{\Delta}- 3\tilde{\delta}$.
If $|\tilde{\delta}|\le 1$, then we may parametrize $\gamma$ as
$\gamma= -\nu\xi/4$, where $|\xi|\ge 1$. In this case, the
coefficient $c$ has the following structure:
$c=\gamma^4\left(1 +2\nu/\gamma^2 +c_2/\gamma^2 +\ldots\right)$,
where $c_2\sim 1$.
Using formula (\ref{lam1}) and expanding $\sqrt{c}$ as
$\sqrt{c}= \gamma^2\left[1 +\nu/\gamma^2 +c_2/(2\gamma^2)
-\nu^2/(2\gamma^4) +\ldots\right]$, we find, neglecting corrections
of the order of $\nu^{-1}$,
\be
\lambda_+^2/\ep^2 = 1- \tilde{\delta}^2 +
4(\tilde{\delta}\xi -1)/\xi^2.
\label{lamplus0}
\ee
If $|\xi| \gg 1$, then we have the same expression for $\lambda_+$
as in the case of a single resonance mode \cite{D95,DK96}:
$\lambda_+= \sqrt{\ep^2-\delta^2}$. This means that the second mode
goes out of resonance, if its detuning $\Delta$ satisfies the inequality
$|\Delta|\gg \nu$. It is interesting, however, that adjusting two
detunings, one can achieve the maximal possible value
$\lambda_+^{(max)}\approx\ep$ even for $\delta\sim\ep$, when the photon
generation becomes impossible in the single-mode case: this happens for
$\xi_*=2/\tilde{\delta}$, i.e., $\gamma_* =-\nu/(2\tilde{\delta})$.

For $|\tilde{\delta}| >1$ we write
$\gamma =-\,\frac12 \nu/(\tilde{\delta} +\chi)$.
Then the resonance exists for $-1 <\chi < 1$.
Under the condition $\nu\gg 1$, the real eigenvalue $\lambda_+$
can be written as (we neglect corrections of the relative order of
$\nu^{-1}$)
\be
\lambda_{+} \approx
\frac{\ep\,\nu\, \sqrt{1-\chi^2}}{\nu +2\tilde{\delta}^2} .
\label{lamplus}
\ee
Again, one can achieve the maximal possible value
$\lambda_+^{(max)}\approx\ep$,
if $\nu \gg \tilde{\delta}^2$, but for $\nu \ll \tilde{\delta}^2$
the maximal eigenvalue decreases as $\ep\,\nu/(2\tilde{\delta}^2)$.

The resonance width $\Gamma_{\Delta}$ (the distance between two points
of intersections between the vertical line $\tilde{\delta}=const$
and the hyperbolas limiting the region of resonance in figure 4) equals
$\Gamma_{\Delta}=\nu/(\tilde{\delta}^2 -1)$ (for $|\tilde{\delta}|> 1$),
so it decreases rapidly for $|\tilde{\delta}|\gg 1$. Similar widths
$\Gamma_{\delta}^{(l,r)}$ (determined in an obvious way from the points
of intersections between the horizontal line $\tilde{\Delta}=const$
and the hyperbolas) are given by the expressions
$\Gamma_{\delta}^{(l,r)}=1 \pm \sigma$, where
\[
\sigma=\frac16\left[\sqrt{(\tilde{\Delta}+3)^2 +6\nu}
-\sqrt{(\tilde{\Delta}-3)^2 +6\nu}\right].
\]
The sum of these widths does not depend on $\tilde{\Delta}$, and
the smallest of them decreases as
$\Gamma_{\delta}^{(small)}\approx 3\nu/(4\tilde{\Delta}^2)$ for
$|\tilde{\Delta}|\gg\nu$.

Simple explicit solutions of the equations of motion can be found in the
special case $\tilde{\delta}=1$, $\gamma=-\nu/2$ (when the generation stops
for the single mode, but the increment takes its maximal value for coupled
two modes). In this case, the eigenvalues of matrix ${\cal A}$ (\ref{matrA})
are as follows,
\[
\lambda_+=R= 1-\frac{2}{\nu} +{\cal O}(\nu^{-2}),
\]
\[
\lambda_-= iJ, \quad J= \frac{\nu}{2} +1 +{\cal O}(\nu^{-2}).
\]
With an accuracy up to terms of the order of $\nu^{-1}$,
the solutions read as
\beqn
&&x_1(t) = x_1(0)
\left[\left(1-\frac{2}{\nu}\right) C_{1}^{-}(2R\tau;t)
+\frac{2}{\nu}\cos\phi_1 \right]
\nonumber\\&&
- p_1(0)
\left[\left(1-\frac{2}{\nu}\right) S_{1}^{-}(2R\tau;t)
-\frac{2}{\nu}\sin\phi_1 \right]
\nonumber\\&&
+\frac{x_3(0)}{4\mu}
\left[C_{1}^{-}(2R\tau;t) -\cos\phi_1 \right]
\nonumber\\&&
-\frac{p_3(0)}{12\mu}
\left[S_{1}^{-}(2R\tau;t) +\sin\phi_1 \right]
\label{x1t1}
\eeqn
\beqn
&&x_3(t) = x_3(0)
\left[\left(1-\frac{2}{\nu}\right)\cos\phi_3
+\frac{2}{\nu} C_{3}^{-}(2R\tau;t) \right]
\nonumber\\&&
+\frac13 p_3(0)
\left[\left(1-\frac{2}{\nu}\right)\sin\phi_3
-\frac{2}{\nu} S_{3}^{-}(2R\tau;t) \right]
\nonumber\\&&
+\frac{x_1(0)}{12\mu}
\left[C_{3}^{-}(2R\tau;t) -\cos\phi_3 \right]
\nonumber\\&&
-\frac{p_1(0)}{12\mu}
\left[S_{3}^{-}(2R\tau;t) +\sin\phi_3 \right],
\label{x2t1}
\eeqn
where the notation is the same as in the preceding section, and
\[
\phi_{k}(\tau;t)=k\overline{\omega} t - 2J\tau.
\]

The mean energies of each mode depend on time as follows
(we neglect corrections
of the order of $\nu^{-2}$, because they are small; moreover, they
increase with time slower than the preserved terms of the order of
$\nu^{-1}$):
\beqn
{\cal E}_1 &=& \frac{\theta_1}{2}\left[
\left(1-\frac{4}{\nu}\right)\cosh(4R\tau) + \frac{4}{\nu}\psi(\tau)
\right]
\nonumber\\&& +
\frac{\theta_3}{\nu}\left[\cosh(4R\tau) + 1 -2\psi(\tau)\right],
\label{E1ofr}
\eeqn
\beqn
{\cal E}_3 &=& \frac32{\theta_3}\left[
1-\frac{4}{\nu} + \frac{4}{\nu}\psi(\tau)\right]
\nonumber\\&& +
\frac{3\theta_1}{\nu}\left[\cosh(4R\tau) + 1 -2\psi(\tau)\right],
\label{E3ofr}
\eeqn
where
\[
\psi(\tau) \equiv \cosh(2R\tau)\cos(2J\tau).
\]
We see that the energy of the first mode is almost the same
that it would be in the case of a single mode under the condition
of the strict resonance, i.e., $\theta_1\cosh(4\tau)/2$.
The energy of the third mode is significantly less than the energy of the
first mode, if $\nu\gg 1$.
Therefore this regime of excitation can be named ``asymmetrical''.
For $\tau>1$, ${\cal E}_3/ {\cal E}_1\approx 6/\nu$.
Note, however, that for the cubical
cavity with $\nu=50/3$, the energy of the third mode is only three times
less than that of the first one.
It is important, nonetheless, that
the rates of increase of the energies of each mode are
almost twice bigger than they were in the case of the strict resonance
discussed in the preceding section.

For the invariant uncertainty products we obtain, with the same accuracy,
almost identical expressions:
\beqn
{\cal D}_{1,3} &=& \frac{\theta_{1,3}^2}{4}\left[
1-\frac{8}{\nu} + \frac{8}{\nu}\psi(\tau)\right]
\nonumber\\&& +
\frac{\theta_1\theta_3}{\nu}\left[\cosh(4R\tau) + 1 -2\psi(\tau)\right].
\label{IUP1}
\eeqn
We see a drastic difference from the strict resonance case: now the
IUP of each mode increases (asymptotically) exponentially with time.
For $\tau\gg 1$, each mode appears in a highly mixed
quantum state, with
${\cal D}_{1}= {\cal D}_{3}=\theta_1\theta_3\exp(4R\tau)/(2\nu)$.

The photon distributions in the first and the
third modes turn out essentially different (and different from the
distributions in the strict resonance case). We have shown in the
preceding section, that the form of the PDF (oscillating or not) depends
on the ratio $\tilde{\cal E}/{\cal D}$. For the first mode,
$\tilde{\cal E}_1/{\cal D}_1 \approx \nu/(2\theta_3)$ for $\tau\gg 1$.
Consequently, the asymptotical PDF oscillates for the initial vacuum
state and for initial thermal states with not very large mean numbers
of photons (less than, approximately, $\nu/4$). Also, the squeezing
coefficient (\ref{s}) tends to the finite asymptotical value
$s_1^{\infty}=
\lim_{\tau\to \infty}({\cal D}_1/\tilde{\cal E}_1) = 2\theta_3/\nu$
(typical dependences $s_1(\tau)$ are given in figure~5).
This resembles (qualitatively) the behavior of the squeezing
coefficient in the one-dimensional cavity with an infinite number
of coupled modes \cite{DA}.
On the contrary, since ${\cal D}_3/\tilde{\cal E}_3 \to \theta_3$ for
$\tau\to \infty$, the final squeezing coefficient of the third mode
coincides with the initial value:
$s_3(0)=s_3^{\infty}= \theta_3 \ge 1$, with very small deviations from this
value in the interval $0<\tau<\infty$.
The photon distribution function of the third mode does not oscillate,
being close to the ``quasigeometric'' distribution (\ref{dist-mon}).

\section{Conclusion}

We have studied the problem of photon creation due to the nonstationary
(dynamical) Casimir effect in a three- or
two-dimensional ideal cavity with an oscillating boundary, in the case when
the spectrum of eigenfrequencies contains two frequencies, whose
ratio is close to $3$, and the boundary oscillates at the frequency
close to the double lowest field eigenfrequency. We have calculated the mean
energy, squeezing coefficient, invariant uncertainty product, and the
photon distribution in each of two resonance modes, both under the strict
and approximate resonance conditions. We have shown that,
due to a strong coupling between the resonance modes, it is possible
to obtain the photon generation even for relatively large detunings between
the frequencies of the wall and the field,
compensating one detuning by another.
However, the resonance width decreases with increase of
detuning, as well as the increment of the exponential growth of the energy.
We have considered both the vacuum and thermal initial states, having
demonstrated not only an amplification of the number of created photons
due to the initial thermal fluctuations, but also such effects as the
``purity exchange'' between the modes and the transformation of the
initial smooth photon distribution to an oscillating distribution,
typical for strongly squeezed states.
        Our results show that choosing a proper set of parameters
of realistic three-dimensional cavities,
one could facilitate observing
the nonstationary Casimir effect, which is a challenge for
experimentalists. However, a correct account of the influence of
damping is still an unsolved problem.

\section*{Acknowledgement}
The authors are grateful to the Brazilian agency CNPq for the support.

%\end{multicols}
%\vfill
\newpage

%\newpage

\begin{figure}
\caption{
Mean energies (normalized by their initial values at $\tau=0$)
of the first and the third modes of the cubic cavity with $\nu=50/3$
in the case of the strict resonance, versus the dimensionless
``slow time'' $\tau$ (\ref{deftaurho}). The order of the curves
in the interval of time between 1.4 and 1.6 is as follows (from bottom
to top):
the first mode for $\theta_{31}=1/3$
(high-temperature initial states);
the first mode for $\theta_1=\theta_3=1$ (vacuum initial state);
the third mode for $\theta_1=\theta_3=1$;
the third mode for $\theta_{13}=3$.
}
\end{figure}

\begin{figure}
\caption{
The photon distribution function ${\cal P}(n)$ for the first mode
and the initial vacuum state, under the condition of the strict resonance,
for the ``slow time'' $\tau=5\pi/(2\rho)$ and $\nu=50/3$.
}
\end{figure}

\begin{figure}
\caption{
The photon distribution function ${\cal P}(n)$ for the first mode
and the initial thermal state with $\theta_1=5$
(i.e., for $\langle n(0)\rangle= 2$ in the first mode),
under the condition of the strict resonance,
for the ``slow time'' $\tau=5\pi/(2\rho)$ and $\nu=50/3$.
}
\end{figure}

\begin{figure}
\caption{
Regions of photon generation in the plane $(\tilde{\delta},\tilde{\Delta})$
of the detuning parameters (tildes have been omitted).
The region of ``symmetrical generation''
$|\tilde{\Delta} -4\tilde{\delta}|<\eta_c$
is between two dashed (almost straight) lines.
Two regions of ``asymmetrical generation'' are confined with
hyperbolas drawn as continuous curves.
}
\end{figure}

\begin{figure}
\caption{
The squeezing coefficient (\ref{s}) versus ``slow time'' $\tau$
(\ref{deftaurho}) for the first mode in the case of ``asymmetrical
generation'', for $\tilde{\delta}=1$, $\gamma=-\nu/2$, and
$\nu=50/3$. The lower curve corresponds to the case $\theta_1=1$
(vacuum initial state). The upper curve corresponds to the case
$\theta_1=5$.
}
\end{figure}

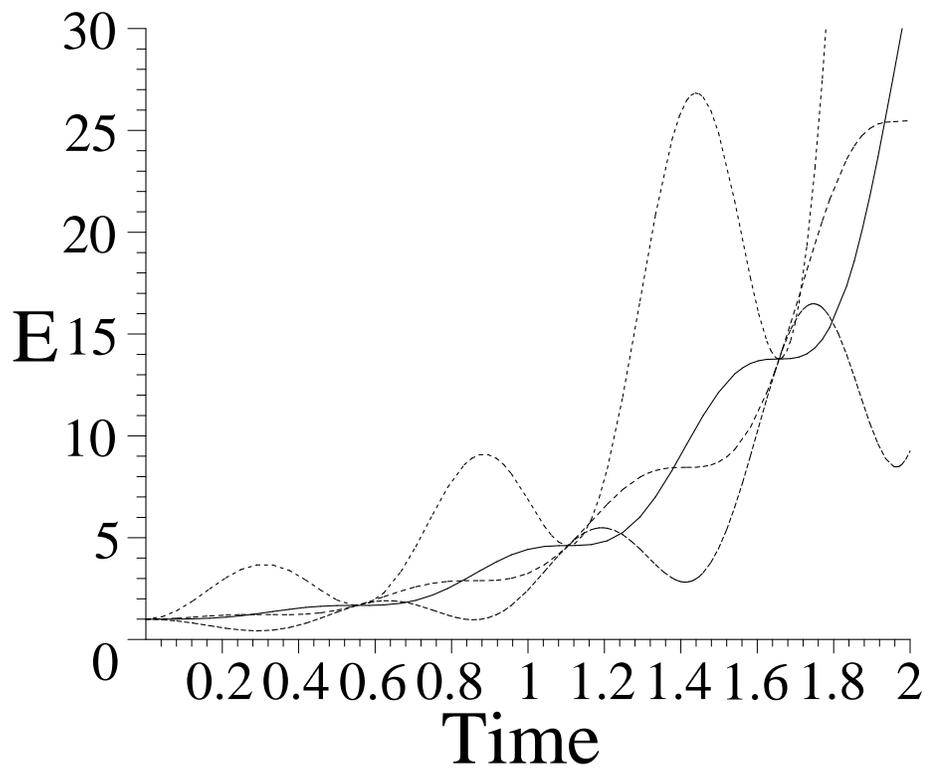
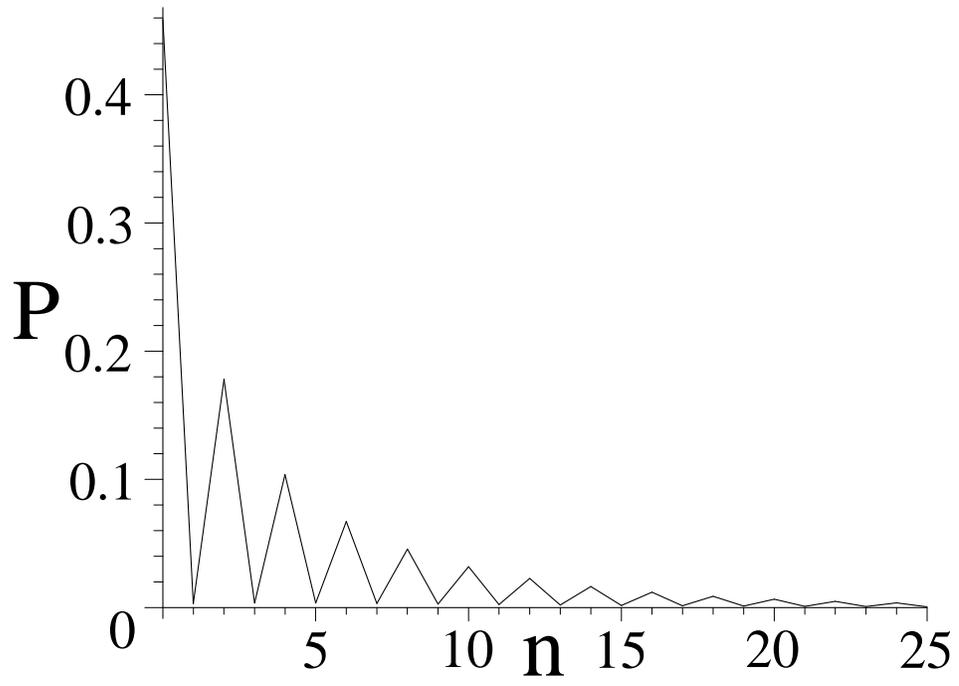
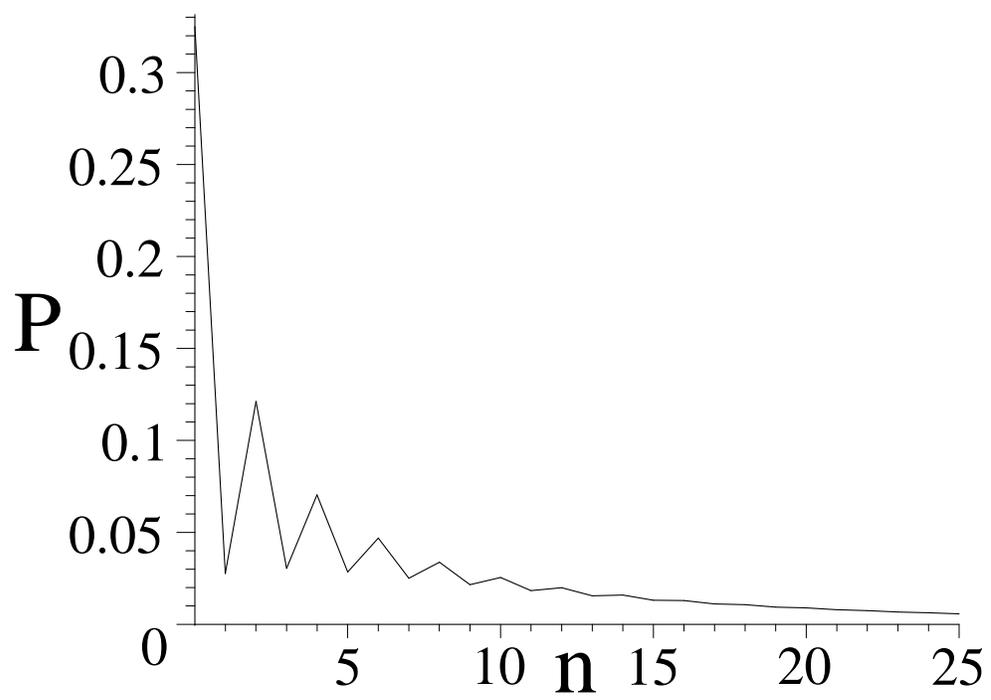
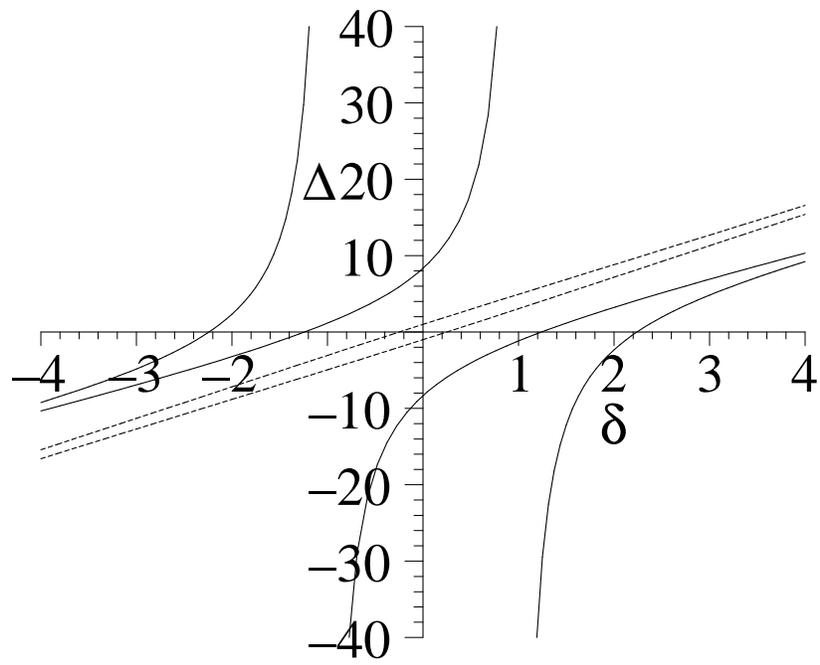
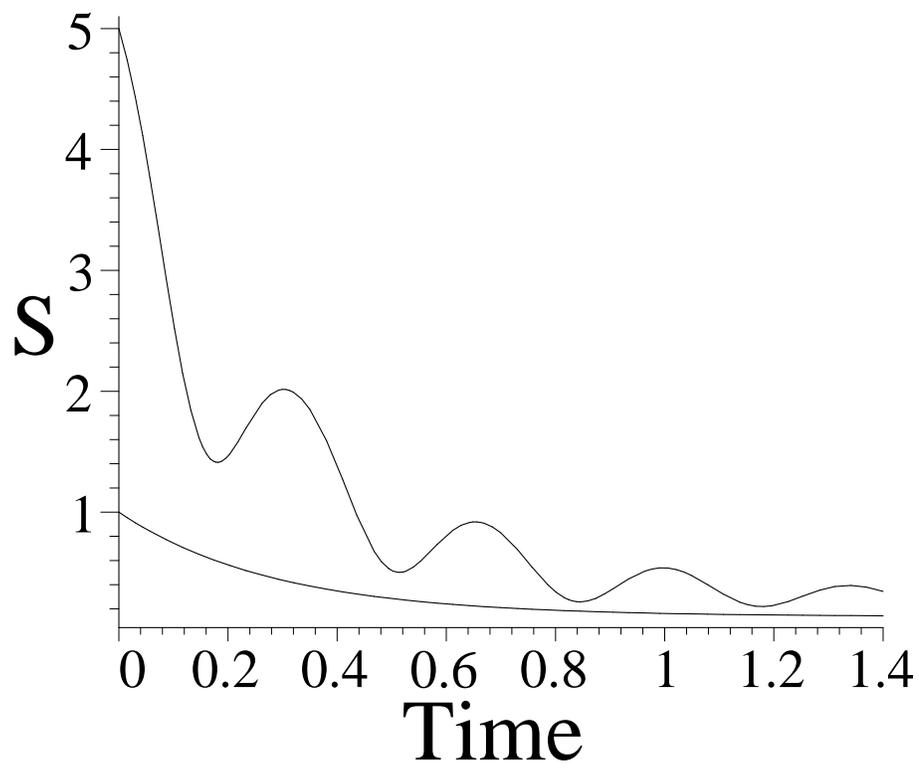
\end{document}